# Modeling and design of a BES diagnostic for the negative ion source NIO1[a]

M. Barbisan[1,b], B. Zaniol[1], M. Cavenago[2] and R. Pasqualotto[1]

[1]*Consorzio RFX - Associazione Euratom-Enea sulla Fusione, corso Stati Uniti 4, I-35127 Padova, Italy*
[2]*INFN-LNL, v.le dell'Università 2, 35020 Legnaro (PD), Italy*

Consorzio RFX and INFN-LNL are building a flexible small ion source (NIO1) capable of producing about 130 mA of H$^-$ ions accelerated at 60 KeV. Aim of the experiment is to test and develop the instrumentation for SPIDER and MITICA, the prototypes respectively of the negative ion sources and of the whole neutral beam injectors which will operate in the ITER experiment. As SPIDER and MITICA, NIO1 will be monitored with Beam Emission Spectroscopy (BES), a non–invasive diagnostic based on the analysis of the spectrum of the H$\alpha$ emission produced by the interaction of the energetic ions with the background gas. Aim of BES is to monitor direction, divergence and uniformity of the ion beam. The precision of these measurements depends on a number of factors related to the physics of production and acceleration of the negative ions, to the geometry of the beam and to the collection optics. These elements were considered in a set of codes developed to identify the configuration of the diagnostic which minimizes the measurement errors. The model was already used to design the BES diagnostic for SPIDER and MITICA. The paper presents the model and describes its application to design the BES diagnostic in NIO1.

## I. INTRODUCTION

Future fusion experiments, like ITER, will require the use of beams of high energy (1 MeV) H/D particles to heat the plasma. At the moment the most reliable solution consists in the neutralization of large fluxes of H$^-$/D$^-$ particles created in an RF plasma source coupled to a system of acceleration grids. This technique has been extensively studied and now Consorzio RFX is going to host the prototypes of the ion source (SPIDER) and of the whole neutral beam injector (MITICA) which will be installed in ITER[1]. In addition to these facilities, a smaller RF negative ion source was built in collaboration with INFN-LNL: NIO1 (Negative Ion Optimization 1). This source generates 9 beamlets of H$^-$ ions accelerated at 60 keV for a total current of 130 mA, and is properly cooled for long time operation[2]. Aim of the experiment is to provide a test facility for the materials, the instrumentation and the numerical simulations for SPIDER and MITICA; the modular design of the machine allows to study the physics of the source with a large freedom in the configuration of the experiment. Similarly to SPIDER[3], NIO1 will be provided with a large number of diagnostics[4]: besides the electric and water cooling plant related measurements of current, pressure, flow and temperature, there will be emission spectroscopy and cavity ring-down and laser absorption spectroscopy to monitor the plasma in the source. The accelerated beam will be analyzed with a fast emittance scanner and its intensity profile with visible tomography, while its power distribution on the calorimeter will be monitored by an infrared camera. An important beam diagnostic is beam emission spectroscopy (BES), which measures divergence, uniformity and direction of propagation of the beamlets, as well as the fraction of ions neutralized inside the system of grids. This does not perturb the beam continuity of operation and therefore it can be used over long beam pulses. It will be also a benchmark of the BES systems to be installed in SPIDER and MITICA[5], therefore its design and components should be as much compatible as possible. After a brief recall of the principles of BES, the paper will show how the diagnostic design has been optimized in order to minimize the measurement errors and, finally, the foreseen diagnostic performances will be reported.

## II. BEAM EMISSION SPECTROSCOPY

Beam Emission Spectroscopy is based on the interaction of the energetic ions with the molecules of the background gas present in the vacuum chamber, in front of the ion source grids. By collision with the molecules, excited neutral particles are produced with the subsequent emission of photons. The most intense emitted radiation corresponds to H$_\alpha$/D$_\alpha$ (n=3 to n=2) transition. In the frame of reference of the laboratory, the wavelength of this spectral line is Doppler shifted according to the following formula:

$$\lambda' = \lambda_0 \frac{1 - \beta \cos\alpha}{\sqrt{1 - \beta^2}} \quad (1)$$

where $\lambda'$ and $\lambda_0$ are the observed and the unshifted H$\alpha$ wavelength (656.2 nm), $\beta$ is the ratio between the speed of the ions and the speed of light (1.13·10$^{-2}$ for 60 keV) and $\alpha$

---

is the angle between the de-excited neutral trajectory and the observation direction. In NIO1 the beamlets are arranged in a square matrix of 3x3 and the beam emission will be collected along three horizontal and three vertical lines of sight (LOS), aligned on the beamlet axes. For each LOS the light is collected by a lens and conveyed onto an optical fiber connected to a grating spectrometer. All 6 fibers are piled along the entrance slit of the spectrometer, so that the resulting spectra can be acquired simultaneously with a 2D CCD camera. Several beam parameters can be deduced from the analysis of each spectrum. Since Hα radiation with no Doppler shift is always present (Figure 3), it's easy to measure the wavelength separation between the shifted and unshifted Hα components; this allows to calculate α (i.e. the orientation of the beamlets) exploiting equation 1 (β is known from the grid voltages). The linewidth $\Delta\lambda$ of the shifted Hα component is determined by the quadratic sum of a number of broadening effects[6]: the intrinsic width of the line $\Delta\lambda_I$, the voltage ripple $\upsilon$ of the grids, the broadening introduced by the spectrometer instrumental function $\Delta\lambda_N$, the angle ω with which the lens is seen by the emitting particle (optical aperture) and the (average) divergence ε of the beamlets:

$$\Delta\lambda = \sqrt{\Delta\lambda_I^2 + \Delta\lambda_N^2 + \left(\frac{\lambda_0}{\sqrt{1-\beta^2}}\beta\sin\alpha\right)^2(\omega^2+\varepsilon^2) + \left(\frac{e\lambda_0}{mc^2\beta}(\beta-\cos\alpha)\right)^2\upsilon^2} \quad (2)$$

where $m$ is the mass of the H/D atom. The divergence of the beamlets can then be obtained from the measurement of $\Delta\lambda$ in the spectra and from the estimation of the other broadening factors. The uniformity of the beam can be measured comparing the integral of the Doppler shifted component from the spectra of different LOS; the ratio between the line integrals however depends on the intersection between beamlets and LOS, and then on their relative alignment.

## III. OPTIMIZATION OF THE DIAGNOSTIC

The BES diagnostic in NIO1, being also a prototype of the similar systems in SPIDER[5] and MITICA, has to fulfill the same requests in terms of accuracy of the measurement of uniformity and divergence, i.e. 10% for both[5]. In particular, the aspects which drive the optimization of the diagnostic are related to the divergence measurement. The error on the measured divergence depends on the accuracy of the quantities in equation 2 ($\Delta\lambda$ included) and on their magnitude. If the contribution of the divergence to the line broadening were too low compared to the other contributions, then a measure of ε with enough precision would not be possible. In the design of NIO1 BES diagnostic a set of parameters (Table I) has been fixed because of the geometry of the experiment or because of the available instrumentation. For example, the CCD camera is the same to be used also in SPIDER[5] and MITICA (1024x1024 pixels, cooled CCD), while the available spectrometer is of Czerny-Turner type, with a grating of 1200 grooves/mm and a focal length of 500 mm. The properties of camera and spectrometer determine $\Delta\lambda_N$ and the plate factor, which is relevant for the measurement of both the centroid and the width of the Doppler shifted line.

**Table I:** Fixed parameters relevant to the estimation of the divergence precision

| Quantity | value | error |
|---|---|---|
| Divergence (e-folding) | 5÷10 mrad (7 mrad as typical) | - |
| Typical error in the measurement of $\Delta\lambda$ | 0.05 pixel | - |
| Plate factor of the spect. $d$ | ~1.8·10$^{-2}$ nm/pix | ~10$^{-5}$ nm/pix |
| Typical entrance slit width of the spect. | 100 μm | - |
| Pixel size of CCD | 13 μm | negligible |
| Typical distance of beamlets from the lens $L_{los}$ | 30 cm | - |
| Acceleration Voltage | 60 kV | 1% |
| Voltage ripple | 0.2 % | negligible |
| Error of ω due to width and different position of beamlets | 7.1 % | - |

At this point, the only diagnostic parameters left to vary are ω and α. The accuracy of divergence estimate depends on them as shown in Figure 1.

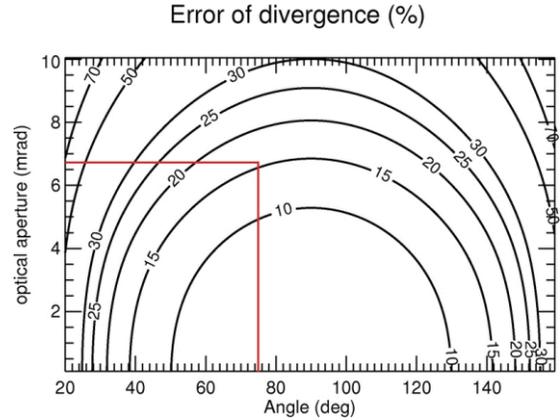

**Figure 1:** Dependence of the divergence accuracy on ω and on the angle of observation, assuming *ε=7 mrad*.

Lowering the angle of observation would worsen divergence accuracy, while increasing it would bring the shifted peak too much near the unshifted $H_a$ line. Moreover, ω cannot be set too low otherwise the reduction of the signal intensity and therefore of its SNR would spoil the measurement of $\Delta\lambda$. The best compromise was found adopting a viewing angle α of 75° as in SPIDER BES[5], and designing the optical head such to keep an ω not over 6÷7 mrad. Due to the further requirements of keeping the AxΩ constant along the LOS and of using optical fibers of *400 μm* core diameter, a lens with clear aperture diameter $d_{lens}$= 4 mm and focal length *f=50 mm* is used. When focused at a distance of 550 mm (far enough beyond the intercept with the beam) the lens gives an almost constant throughput *AxΩ* of 5.22·10$^{-10}$ m$^2$ srad along the LOS, see Figure 2. With these settings, the relative error for the divergence estimate is about 15 % for *ε=7 mrad*; in general, the accuracy varies in the expected divergence range from 7% (ε=10 mrad) to 30% (ε=5 mrad). Though the request of 10% is presently not always satisfied, this is the best that can be achieved with the available instrumentation. An improvement could be obtained reducing ω (but at the expense of the signal strength) or the plate factor of the spectrometer.

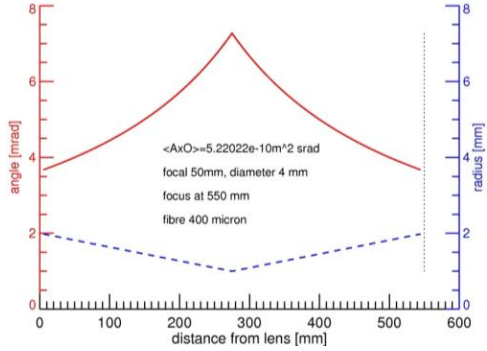

**Figure 2:** LOS radius (dashed line) and effective optical aperture variation (solid line) along the LOS.

## IV. SIMULATION AND ANALYSIS OF BES SPECTRA

Once the essential parameters of the diagnostic were chosen, a set of codes was developed to simulate the spectra obtainable with the adopted set-up. For this aim, the ion beamlets were modeled with a gaussian transversal profile:

$$n(\delta,l) = \frac{\bar{n}_G r_G^2}{2\sigma_b^2(l)} e^{-\frac{\delta^2}{2\sigma_b^2(l)}} \quad (3)$$

$n_G$ is the mean density of ions at the hole of radius $r_G$ in the last grid, $\delta$ is the distance from the beamlet axis and $\sigma_b(l)$ is the gaussian width of the beamlet at a distance $l$ along its axis:

$$\sigma_b(l) = \sigma_0 + l \cdot \tan\left(\frac{\varepsilon}{\sqrt{2}}\right) \quad (4)$$

where $\sigma_0$ is the effective radius of the beamlet at the last grid; from a comparison between the grids of SPIDER and NIO1 it can be assumed that $\sigma_0 = 2$ mm. For the stripping losses, the same model was adopted, assuming $\varepsilon=30$ mrad as in SPIDER. The simulation identifies the volume of the beamlets which is effectively seen by the considered LOS and integrates the emissions assuming that the beamlets interact with a gas of hydrogen at a uniform pressure 0.05Pa and temperature 300 °K. Each contribution is added to the spectrum as a gaussian curve with centroid and width calculated respectively from equations 1 and 2. The conversion from the amount of collected photons to CCD counts is performed assuming a CCD exposure time of 100 ms and an overall loss factor of ~70 % which accounts for the optical losses of the instrumentation and for the efficiency of the CCD camera and the spectrometer. The simulated spectrum for a vertical LOS is shown in Figure 3. The small peak between the main two is produced by the losses of ions in the acceleration stage. Spectra were produced for different values of divergence and then analyzed to test the measurement of ε in realistic conditions; the results validate the expected relative error of ε. Moreover the precision of the uniformity measurements was tested with different extraction densities and slightly varying position and direction of the LOSs: the relative error is below the requested 10 %, however to avoid systematic errors larger than this threshold the optics should be aligned within ±1 mm of offset and ±3 mrad of tilting.

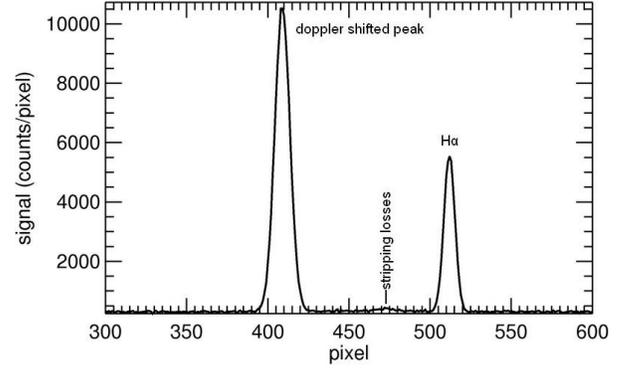

**Figure 3:** Simulation of the spectrum collected by BES for a vertical LOS.

## V. CONCLUSIONS

The BES diagnostic for NIO1 was optimized to maximize the precision in the measurement of the desired quantities. With the available instrumentation, only the error of the divergence should exceed the threshold of 10%, but still within acceptable limits. A more accurate simulation of the BES behavior validated the estimates and allowed the evaluation of the alignment precision for the optics which must be respected during the installation of the diagnostic.

## ACKNOWLEDGMENTS

This work was set up in collaboration and financial support of Fusion for Energy and INFN.